\documentclass[10pt]{article}

\usepackage[dvips]{graphicx,rotating}
\usepackage{subfigure}
\usepackage[]{amsmath}

\begin{document}

\noindent{\large Reliability of Tumour Classification from Multi-Dimensional DCE-MRI Variables using Data Transformations.}

\vspace{0.2in}

\noindent S. V. Notley$^1$, N. A. Thacker$^2$, L. Horsley$^3$, R. A. Little$^2$, Y. Watson$^2$, S. Mullamitha$^3$, G. C. Jayson$^3$ and A. Jackson$^2$.

\vspace{0.2in}

\noindent{\small [1] Department of Automatic Control and Systems Engineering, University of Sheffield, Sheffield, UK.}

\noindent{\small [2] Division of Informatics, Imaging and Data Science, University of Manchester, Manchester, UK.}

\noindent{\small [3] Department of Medical Oncology, The Christie NHS Foundation Trust, Wilmslow Road, Manchester M20 4BX, UK.}

\noindent {\small \textbf{Contact:} s.v.notley@sheffield.ac.uk}

\subsection*{Abstract}
{\small Summary mean DCE-MRI variables show a clear dependency between signal and noise variance, which can be shown to reduce the effectiveness of difference assessments. Appropriate transformation of these variables supports statistically efficient and robust comparisons. The capabilities of DCE-MRI based descriptions of hepatic colorectal tumour classification was assessed, with regard to their potential for use as imaging biomarkers. Four DCE-MRI parameters were extracted from 102 selected tumour regions. A multi-dimensional statistical distance metric was assessed for the challenging task of comparing intra- and inter- subject tumour differences. Statistical errors were estimated using bootstrap resampling. The potential for tumour classification was assessed via Monte Carlo simulation. Transformation of the variables and fusion into a single chi-squared statistic shows that inter subject variation in hepatic tumours is measurable and significantly greater than intra-subject variation at the group level. However, reliability analysis shows that, at current noise levels, individual tumour assessment is not possible. Appropriate data transforms for DCE-MRI derived parameters produce an improvement in statistical sensitivity compared to conventional approaches. Reliability analysis shows, that even with data transformation, DCI-MRI variables do not currently facilitate good tumour discrimination and a doubling of SNR is needed to support non-trivial levels of classification.}










 
\section*{Introduction}

Knowledge of the gene mutations that drive cancer has led to the development of a large number of mechanism-based therapeutics (MBT). However, there is a clear need to improve trial design to limit patient exposure to ineffective drugs and to accelerate the decision making for new agents. 
Imaging biomarkers are particularly attractive as they allow interrogation of the whole tumour, repeated measurements over time and support studies of inter- and intra-tumoural heterogeneity (\cite{Canuto2009, Rose2009, Fisher2013, Asselin2012, Li2005, Alic2011}). 

Dynamic contrast enhanced MRI (DCE-MRI) has been widely used in clinical trials of new agents and allows estimation of a number of parametric variables describing the tumour vascular micro-environment (\cite{Rose2009}).  Previous studies have described correlations between DCE-MRI characteristics and tumour grade and histological subtype (\cite{Martincich2012, Youk2012}). These are typically weak associations, inadequate for effective patient/tumour stratification even though statistical differences between tumour types may be seen in group comparison studies. There is a clear need to optimise the phenotypic information that we can extract from imaging data to improve the specificity and power of clinical trials using imaging biomarkers and, ideally, to provide sufficient statistical power to support personalised therapeutic decisions in individual patients.  One common approach is the development of image acquisition and analysis strategies to improve the biological specificity and reliability of individual imaging derived parameters. In this work we take a complementary approach, to ensure that the information content of this multi-parametric data is fully leveraged to support decision-making by the use of efficient statistical approaches.

In this study we use hepatic metastatic colorectal cancer as a model to assess the ability of multi-parametric DCE-MRI data to support stratification of phenotypic tumour variation. We show that individual DCE-MRI derived parameters typically have a low information content and are not capable of robustly classifying tumour sub-types with any fidelity. Fusion of the four DCE-MRI parameters into a single discriminating measure, to make better use of the available information, is problematic since the variables are non-commensurate. Further to this the parameters also show a dependency between the variance of the signal measurement noise and the mean value of the variable. 

The use of non-linear transformations to generate variables with uniform independent measurement noise has been described previously (\cite{Bartlett1947, Box1964, Draper1969, Parsons2007, Foi2011, Bland1996a, Bland1996b}). Although this technique has been recently applied to medical images (\cite{Scarpelli2018}) it is not widely recognised in the medical and radiologic literature. The method transforms the variables into a homoscedastic space where measurement noise is independent of the parameters. A common approach, found in the machine learning and pattern recognition literature, attempts to address these issues by scaling variables by their respective ranges or an estimated standard deviation based on the raw variables (\cite{Garcia2009, Gertheiss2009, Altincay2011, Ripley1996}). This does not make any attempt to model the error characteristics of the individual parameters and does not facilitate a conventional statistical difference test. 

In this work we use an approach presented by Notley \emph{et al} (\cite{Notley2023}) that stabilizes parameter noise estimates, based on difference of repeats, transforming variables to a homoscedastic space. This allows the estimation of appropriate summary statistics (\cite{Bland1996a}), combination of non-commensurate variables and robust identification of outliers. This allows a multi-dimensional approach, based upon the construction of a chi-squared statistic, supporting the fusion of information from multiple parameters. The method is evaluated on whole tumour data sets in the analysis of inter and intra-subject variation. Detecting such differences is more challenging than the more common task of detecting differences between normal tissue and pathology, but in line with the use of pharmacokinetic data for the assessment of whole tumour heterogeneity (\cite{Asselin2012}). The evaluation is restricted to varying degrees to demonstrate the general validity of the approach and to avoid drawing false conclusions due to statistical biases in the data.

We hypothesise that: 1) the use of appropriate modelling of the parameter dependent characteristics of measurement error will allow transformation of parametric variables to more closely match the assumptions of standard statistical tests and: 2) that this will result in an increase in discriminatory power over the commonly used analysis approaches. 

Surrogate Monte Carlo datasets for varying numbers of tumour types were constructed based on the observed estimated signal and noise characteristics of the transformed variables. The reliability of correctly classifying tumour types was estimated, for varying signal-to-noise ratios, as a function of classification density (granularity) both with and without variable transformations. We believe that these results are indicative of the ability of current DCE-MRI variables to quantify phenotypic whole tumour heterogeneity both within and across subjects and conclude that to achieve any level of practical consistency the SNR of the these variable needs to be improved.

\section*{Methods}

The data used in this study is from a data set collected as part of a larger drug trial. However, in this work we use the baseline repeatability dataset to investigate the ability of DCE-MRI parameters to discriminate between tumours.

\subsection*{Patient Selection}
The patients included in this study were undergoing imaging
with Dynamic Contrast Enhanced Magnetic Resonance
Imaging (DCE-MRI) as a part of a clinical trial running at our
institution and had given written informed consent
to participate in the study. The study had ethical approval and was carried out in accordance with standards
of GCP. Patients were eligible providing they were over eighteen years of age with biopsy confirmed metastatic
colorectal cancer, without previous therapy for metastatic disease and disease measuring 3cm. All patients had
undergone 2 baseline DCE-MRI scans, median 4 days (range 2-7 days) prior to treatment. It is this data which
has been used in the current work to investigate biological variation within tumour tissues. To help control for
tumour micro-environment, only patients with liver metastases were included in our analyses. This resulted in a
sample of 29 subjects with numbers of metastases varying between 2 and 6, giving 102 tumours in total.

\subsection*{MRI Data Acquisition and Analysis}
Data were acquired on a 1.5T Philips Intera system. The baseline T1 measurement consisted 
of 3 axial spoiled Fast Field Echo (gradient echo) volumes with flip angles 2, 10, 20 degrees, 
respectively and 4 signal averages. The dynamic series was acquired using the scanner whole 
body coil (Q body coil) for transmission and reception. The dynamic series consisted of 
75 consecutively-acquired axial volumes with a flip angle of 20 degrees, 1 signal average, 
and a temporal resolution of 4.97 s. All studies maintained the same number of slices (25), 
field of view (375 mm × 375 mm), matrix size (128 × 128), TR (4.0 ms), and TE (0.82 ms) for 
the baseline T1 measurement images and the dynamic series itself. Slice thickness was 4 mm 
for small target lesions or 8 mm for larger lesions, giving superior-inferior coverage of 
100 mm or 200 mm, respectively. Gadoterate Meglumine (Dotarem®) was injected intravenously 
(IV) by power injector at the time of the sixth dynamic acquisition at 0.2ml/kg, followed 
by a 20ml saline flush at a set rate of 3ml/sec. This was followed by acquisition of a 
post contrast T1-weighted image.

VOIs were delineated by an experienced radiographer on co-registered high resolution 
T1- and T2-weighted images. Whole TV was measured for each lesion. An arterial input 
function was measured where possible; in circumstances where this was not appropriate, 
a population derived input function was used (\cite{Parker2003}). Analysis was performed 
using in-house software (Manchester Dynamic Modelling) and the extended Tofts and Kermode 
pharmacokinetic model(25) was used to calculate the  fractional volume of the extravascular 
extracellular space ($v_e$). The model free measurement, initial area under the gadolinium 
contrast curve at 60s (IAUC60) was calculated and voxels from tumour VOIs were included 
in the analysis if they demonstrated uptake of contrast, this was defined as an initial 
IAUC in the first 60 seconds (IAUC60)$>$0 mmol (See Appendix B for a description related to contrast agent concentration estimation). Where possible, measured Arterial Input Functions (AIFs) were used when data was of sufficient quality
to allow derivation; in other cases a population derived AIF were used.

Median values of the measured parameters $K^{trans}$, $v_p$ and $v_e$ were extracted from
distributions obtained from the 102 selected tumour regions. The enhancing fraction ($E_{frac}$) was also 
measured and then redefined as $E_{frac}'=100(1-E_{frac})$. For each tumour the process was repeated to generate a repeated measures dataset.

\subsection*{Optimal Data Transforms}

In this work we directly consider the characteristics of the repeatability sample noise, in particular heteroscedasticity, which may be determined by comparison of matched pairs. The dependency of the noise on the measurement value may be visualised by plotting the differences of the repeat measurements as a function of the average of the average i.e. Bland-Altman plots. Following the method of Notley \emph{et al} (\cite{Notley2023}) we use a power law transform of the form
\begin{equation}
f(x)=x^\theta
\end{equation}
\noindent that was chosen empirically based upon observation of the Bland-Altman plots. The log-likelihood function was optimised with respect to $\theta$ by exhaustive search over the range -5 to 5.

\subsection*{Inter-Tumour Distance Measures}

The data was first analysed using the standard method of scaling each variable by its standard deviation. A matrix of the measured variables, $\mathbf{V}$, is defined as:

\begin{equation}
\mathbf{V} = 
 \begin{pmatrix}
  k^{trans}(1) & v_p(1) & v_e(1) & E_{frac}'(1) \\
  \vdots  & \vdots  & \vdots & \vdots  \\
  k^{trans}(N) & v_p(N) & v_e(N)& E_{frac}'(N) 
 \end{pmatrix}
\end{equation}

where N is the number of tumours in the dataset ($N=102$). A matrix $\mathbf{R}$ was similarly defined for the repeat measurements.

A distance, $D_{stan}$, between tumours was then defined as:

\begin{equation}
\label{eqn:scale}
D_{stan}^{j,k}=\sum_{i=1}^4 \frac{(\mathbf{V}_{j,i}-\mathbf{R}_{k,i})^2}{\sigma_{\mathbf{V}_{*,i}}^2}
\end{equation}

where $D_{stan}^{j,k}$ is a distance between the $j$-th and the $k$-th tumours in the data set and $\sigma_{\mathbf{V}_{*,i}}$ is the standard deviation of the $i$-th variable ($i$-th column of $\mathbf{V}$) as measured. 

Data transforms, $f(\cdot)$, were estimated and applied to each raw variable to transform the data to the homoscedastic space. The S.D. of the \textit{noise} for the transformed data was then estimated by subtraction of the repeat measurements from the respective initial measurement. A chi-squared variable for 4 degrees of freedom for the measured difference between
tumours $i$ and $j$ was defined as the sum of the squares of the difference between 
changes in each derived DCE MRI variable $i$ divided by its reproducibility variance. 
\begin{equation}
\label{eqn:chi2}
\chi^2_{j,k} = \sum_{i=1}^4 \frac{(f_i(\mathbf{V}_{j,i})-f_i(\mathbf{R}_{k,i}))^2}{\sigma_{\mathbf{n}_i}^2}
\end{equation}

where $f_i(\cdot)$ is the transformation function for the $i$-th variable and $\sigma_{\mathbf{n}_i}$ is the standard deviation of the $i$-th \textit{noise} signal, $\mathbf{n}_i$, a column vector defined as $\mathbf{n}_i=f_i(\mathbf{V}_{*,i})-f_i(\mathbf{R}_{*,i})$, where $\mathbf{V}_{*,i}$ and $\mathbf{R}_{*,i}$ are the $i$-the columns of the matrices $\mathbf{V}$ and $\mathbf{R}$ respectively.

Although equations \ref{eqn:scale} and \ref{eqn:chi2} look similar it is important to note that in equation \ref{eqn:chi2} the denominator is an estimate of the level of \textit{homogeneous noise}. The $\chi^2$ distribution had Poisson like noise characteristics in that the variance of the estimate scales with the value. As discussed above, for statistical efficiency the variable needs to be transformed to a space where the distribution of the distance metric does not change with the value of the metric i.e. the homoscedastic space. In this case the method that gives the homoscedastic space is given by the square root transform (as an approximation to the Anscombe transform). Thus, the quantity $D_{tran}^{j,k}~=~\sqrt{\chi^2_{j,k}/4}$ is expected to have a mean value of 1 for data $j$ and $k$ which differs only due to the presence of the modelled level of measurement error ($\sigma_{\mathbf{n}}$). 
 
The distance measures $D_{stan}$ and $D_{tran}$ were calculated for differences between tumours of each subject, and also for differences between tumours from different subjects. Differences were always taken to the measurement from the alternative baseline study so that the estimated reproducibility was correctly incorporated. Bootstrap resampling (1000 resamples) was used to estimate the mean and standard deviation of these distances. For each bootstrap resampled dataset, the scaling parameters were re-estimated. The statistical distance between group means was then calculated as:
 
 \begin{equation}
 \label{eqn:stat_dist}
 D_{stat}=\sqrt{\frac{(\mu_{inter} - \mu_{intra})^2}{\sigma_{inter}^2+\sigma_{intra}^2}}
 \end{equation}
 
 where $\mu_{inter}$ and $\mu_{intra}$ are the inter and intra group means respectively; $\sigma_{inter}$ and $\sigma_{intra}$ are the respective group standard deviations. For a large number of samples, this statistical distance may be interpreted as a z-score and significance levels (p-values) were calculated from this using the standard integration of the error function. Errors on the statistical distance and z-scores were estimated using error propagation (28).

With such a small dataset, consisting of only 29 subjects, bias may be introduced, especially in the inter-subject distances, due to combinatorial effects of some subjects having  more tumours than others. To reduce this effect we defined a Maximum Number of Tumours Per Subject (MNTSP). Distances were computed for a MNTSP ranging from 2 to 6. This approach was also used to gain insight into the performance gains made with smaller data sets. 

\subsection*{Reliability Analysis}

The reliability of DCE-MRI parameters to stratify tumour variation was investigated by formulating the problem as a classification task based on the multi-dimensional distance measure (see above). Monte Carlo datasets were generated to match the distributions found on the measured signal and noise characteristics. The average correct classification of tumours was assessed at varying levels of categorisation and noise.

For simulation purposes, metastases are assumed to be genetic clones, within each subject, with phenotypic biological variation due to variations in the micro-environment. For each tumour type/class, a clonal 'center', in the transformed variable space, is randomly chosen, from a normal distribution, based on the measured mean and standard deviation of each variable. A ground truth dataset of individual tumours, with biological variation, was generated as random samples, with normal distribution, around each clonal center. 

A full repeated measures dataset was then generated by inclusion of additive, normally distributed, measurement noise to the ground truth dataset. The corresponding heteroscedastic dataset was then generated by applying the inverse of the transform functions estimated from the real dataset.

Surrogate repeated measure datasets were generated with the number of tumour types/classes in the range 2 to 30 for both homoscedastic and heteroscedastic datasets; two thousand surrogate datasets were generated for each. Each tumour of the noisy datasets were classified based on the closest clonal center measured using the statistical distance described above. The average percentage of correct classifications across the two thousand surrogate datasets was computed. The analysis was repeated with the measurement noise halved (the signal-to-noise ratio doubled).

\section*{Results}

\begin{figure}[!ht]

\includegraphics[scale = 0.7]{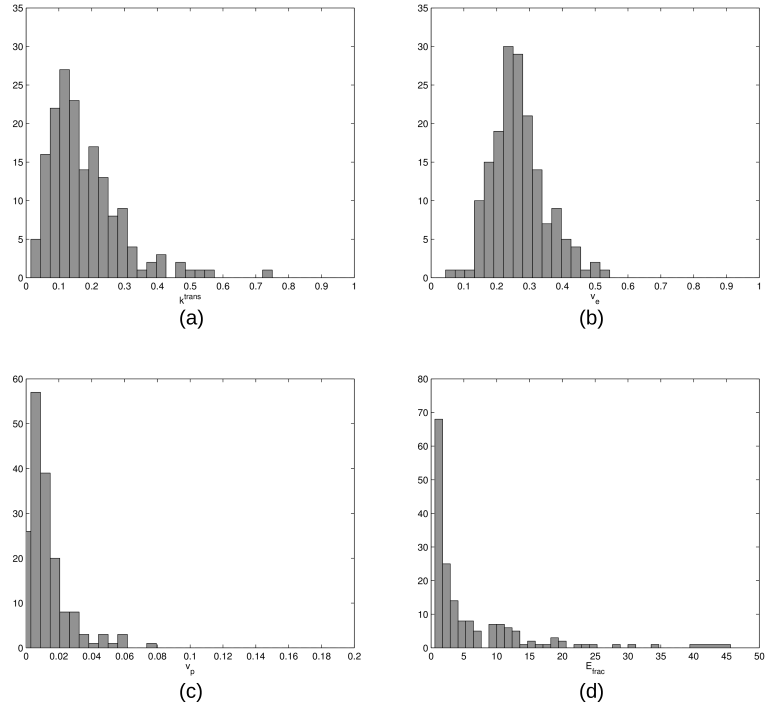}

		\caption{
			{\it
				The distribution of the median DCE MRI variables $K^{trans}$, $v_e$, $v_p$ and
				$100(1- E_{frac})$.
			}
		}\label{fig:col_hist}

\end{figure}

\begin{figure}[!ht]

\includegraphics[scale = 0.7]{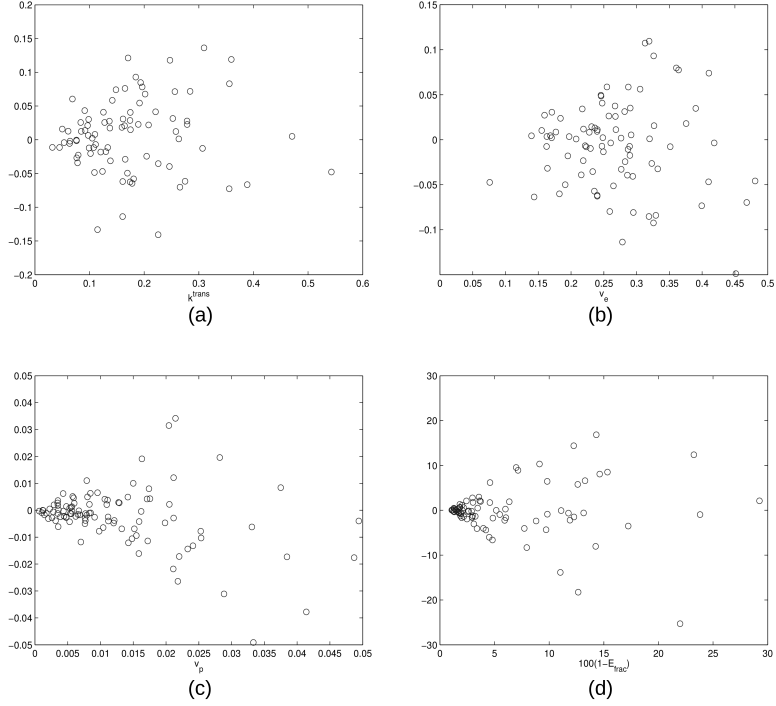}

		\caption{
			{\it 
				Bland-Altman plots, showing reproducibility for median values of 
				$k^{trans}$, $v_p$, $v_e$ and $E_{frac}$.
				A clear parameter dependant accuracy is seen for all variables.
			}\label{fig:col_BA}
		}

\end{figure}

Figure \ref{fig:col_hist} shows histograms  and Bland-Altman plots, figure \ref{fig:col_BA}, constructed from the repeat data show a parameter dependent 
measurement repeatability. As these repeat measurements were obtained from 
different scan sessions we can assume these estimates include all important 
aspects of the variation intrinsic to the process of measurement and consequently the accuracy
with which we can quantify biological change.

Table \ref{tab:D_stan} shows the results of the computation of statistical measures on both the distance measures $D_{stan}$ and $D_{tran}$. The mean and standard deviation of the mean, calculated using bootstrap resampling, are shown for both intra- and inter-subject groups. The z-score and corresponding p-value for the distance between the two groups (equation \ref{eqn:stat_dist}) is also shown. 

\begin{table}[!ht]
	\centering
{\tiny 
	\begin{tabular}{c|c|c|c|c|c|c|c|c}
	   	     & Size &               &                  &               &                                &\\
		 MNTSP & (intra/inter) & $\mu_{intra}$ & $\sigma_{intra}$ & $\mu_{inter}$ & $\sigma_{inter}$ $\sigma$ & Stat. Dist & p-value \\
		\hline 
		$D_{stan}$ & & & & & & & &\\
		\hline
		 2 & 58/59 & 1.54 & 0.738 & 2.96 & 0.488  & 1.6$\pm$0.03 & 0.0548\\
		3 & 75/76 & 1.46 & 0.417 & 2.83 & 0.384  & 2.4$\pm$0.03 & 0.0082\\
		 4 & 85/86 & 1.34 & 0.312 & 2.69 & 0.342  & 3.0$\pm$0.03 & 0.0013\\
		 5 & 92/93 & 1.27 & 0.272 & 2.71 & 0.319  & 3.4$\pm$0.03 & 0.0003\\
		 6 & 96/97 & 1.17 & 0.225 & 2.64 & 0.294  & 3.9$\pm$0.03 & $<$0.0001\\

		\hline
		$D_{tran}$ & & & & & & & &\\
   		\hline
			 2 & 58/59 & 1.41 & 0.193 & 2.27 & 0.173  & 3.3$\pm$0.03 & 0.0004 \\
			 3 & 75/76 & 1.52 & 0.156 & 2.21 & 0.144  & 3.4$\pm$0.03 & 0.0003 \\
			 4 & 85/86 & 1.49 & 0.131 & 2.10 & 0.141  & 3.1$\pm$0.03 & 0.0010 \\
			 5 & 92/93 & 1.44 & 0.118 & 2.09 & 0.136  & 3.6$\pm$0.03 & 0.0001 \\
			 6 & 96/97 & 1.43 & 0.112 & 2.11 & 0.127  & 4.0$\pm$0.03 & $<$0.0001 \\
	\end{tabular}
}
	\caption{\it{Statistical measures made on distance measures $D_{stan}$ and $D_{tran}$}.}
	\label{tab:D_stan}
\end{table}

\subsection*{Standard Distance Measure}

From table \ref{tab:D_stan}, it can be seen that for a small data set with a MNTSP of 2 the standard method fails to find a significant difference between the groups. As the MNTSP is increased the statistical distance between the groups increases and becomes statistically significant.

\subsection*{Homoscedastic Distance Measure}

\begin{figure}[!ht]

\includegraphics[scale = 0.7]{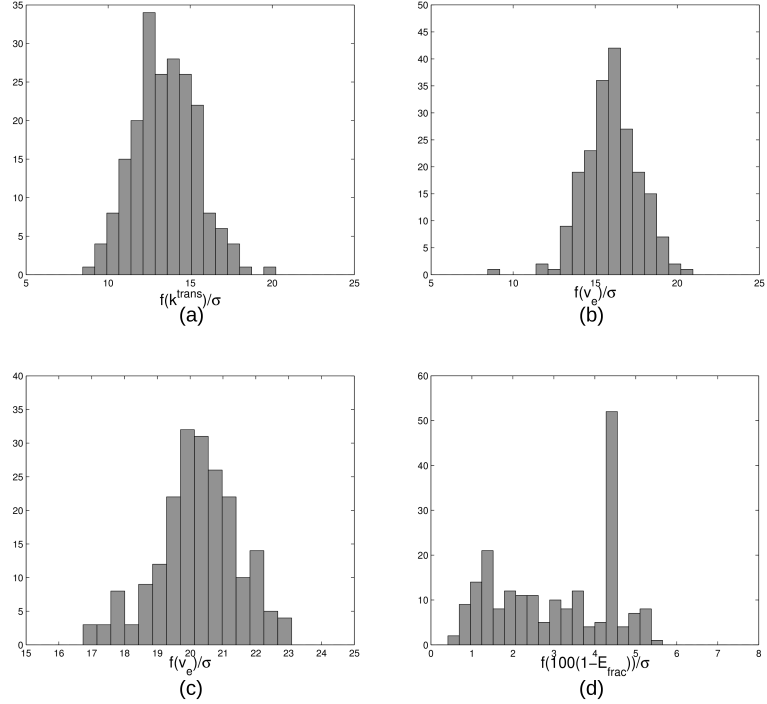}

		\caption{
			{\it
				Transformed variables scaled to reproducibility. Unlike the original
				parameter distributions (Figure 1), the spread of
				each variable (e.g. variance) is a measure of the associated information content.
				Gaussian random variables containing no signal (only noise) are expected to have 
				a Gaussian distribution with unit variance, see Table 2.
			}
		}\label{fig:tHists}

\end{figure}

\begin{figure}[!ht]

\includegraphics[scale = 0.7]{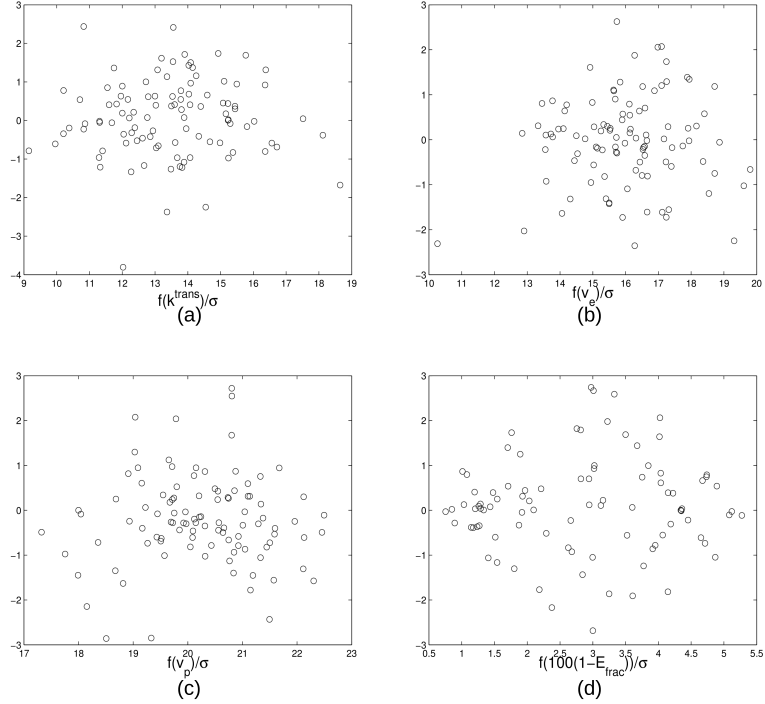}

		\caption{
			{\it
				Bland-Altman plots, showing reproducibility for transformed variables
				derived from $k^{trans}$, $v_p$, $v_e$ and $100(1-E_{frac})$ ,
				scaled on the x axis to units of measured reproducibility $\sigma$.
				For a successful transformation  the residual distributions (distribution of
				scatter above and below zero) should be
				independent of the variable. The high density value in the $100(1-E_{frac})$ plot
				is due to the quantisation of this variable at 100\% which causes
				identical values which cannot be separated by a transformation.
			}
		}\label{fig:tBA}

\end{figure}

The data transform method was applied to each measured variable and the optimal transform values shown in table \ref{tab:trans} were found. Figure \ref{fig:tHists} show the histograms of the transformed variables and figure \ref{fig:tBA} shows the corresponding Bland-Altman plots. These figures show data distributions more closely conforming to a Gaussian and  no evidence of any dependency of repeatability on parameter values.

\begin{figure}[!ht]

\includegraphics[scale = 0.7]{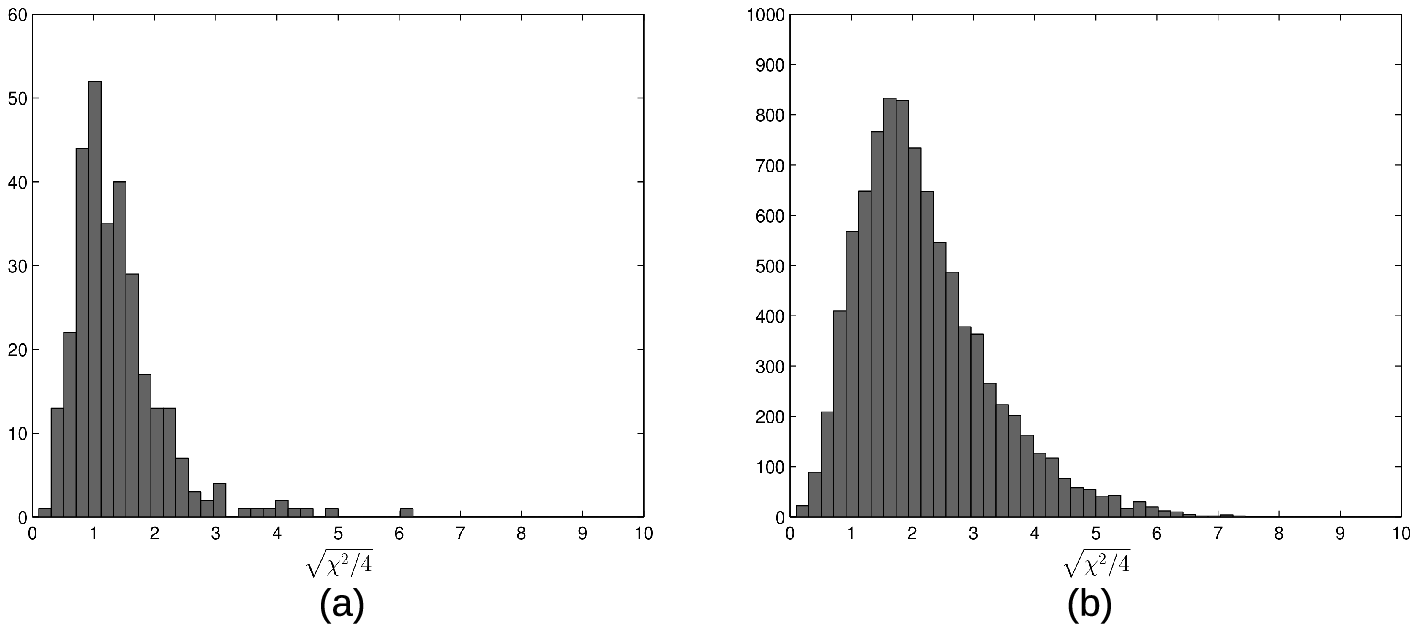}

		\caption{
			{\it
				The distributions of statistical distances, $D_{tran}$, for within subjects (left) and between subject (right) tumours. Computed using a chi squared statistic based on the transformed and scaled DCE MRI summary variables of median $K^{trans}$,$v_{e}$,$v_{p}$ and $100(1-E_{frac})$).
			}}
			\label{fig:distance_distributions}
\end{figure}

The statistical distance, $D_{tran}$, constructed for differences between tumours of each
subject, and also for differences between tumours from different subjects (Figure \ref{fig:distance_distributions}) gave mean values of 1.46 $\pm$ 0.14 and 2.16 $\pm$ 0.18. Both of these values are statistically significantly different to the null hypothesis, that differences are entirely due to measurement noise.
This provides evidence at the population level for not only inter-tumoural heterogeneity  but also for inter-tumoural heterogeneity within single subjects.

Table \ref{tab:trans} shows the standard deviation of each of the transformed variables. The distribution of the transformed and scaled measurements from all 29 subjects were found to have standard deviations between 1.7 and 2.7. As the expected value for pure noise is 1.0, this indicates that the signal pertaining to biological variation for each individual measurement is quite weak, and insufficient to allow effective separation of this tumour data.

The results of the computation of statistical measures on the $D_{tran}$ distance metric are shown in the lower part of table \ref{tab:D_stan}. In this case, application of the transforms results in significant increases in statistical efficiency (Stat. Dist.) over that found using $D_{stan}$.

\begin{table}[!ht]
	\centering
	\begin{tabular}{ccc}
		\hline
		& Transform Parameter $A$  \\
		Variable & ($f(x)=x^A$) & Standard Deviation \\
		\hline
		k-trans & 0.2 & 2.7\\
		$v_{e}$  & 0.3 & 2.4\\
		$v_{p}$ & 0.1 & 1.7\\
		1-$E_{frac}$ & 0.6 & 1.9\\
		\hline
	\end{tabular}
	\caption{\it{Transform parameters for each variable}}
	\label{tab:trans}
\end{table}

\subsection*{Reliability Analysis}

\begin{figure}[!ht]

\includegraphics[scale = 0.7]{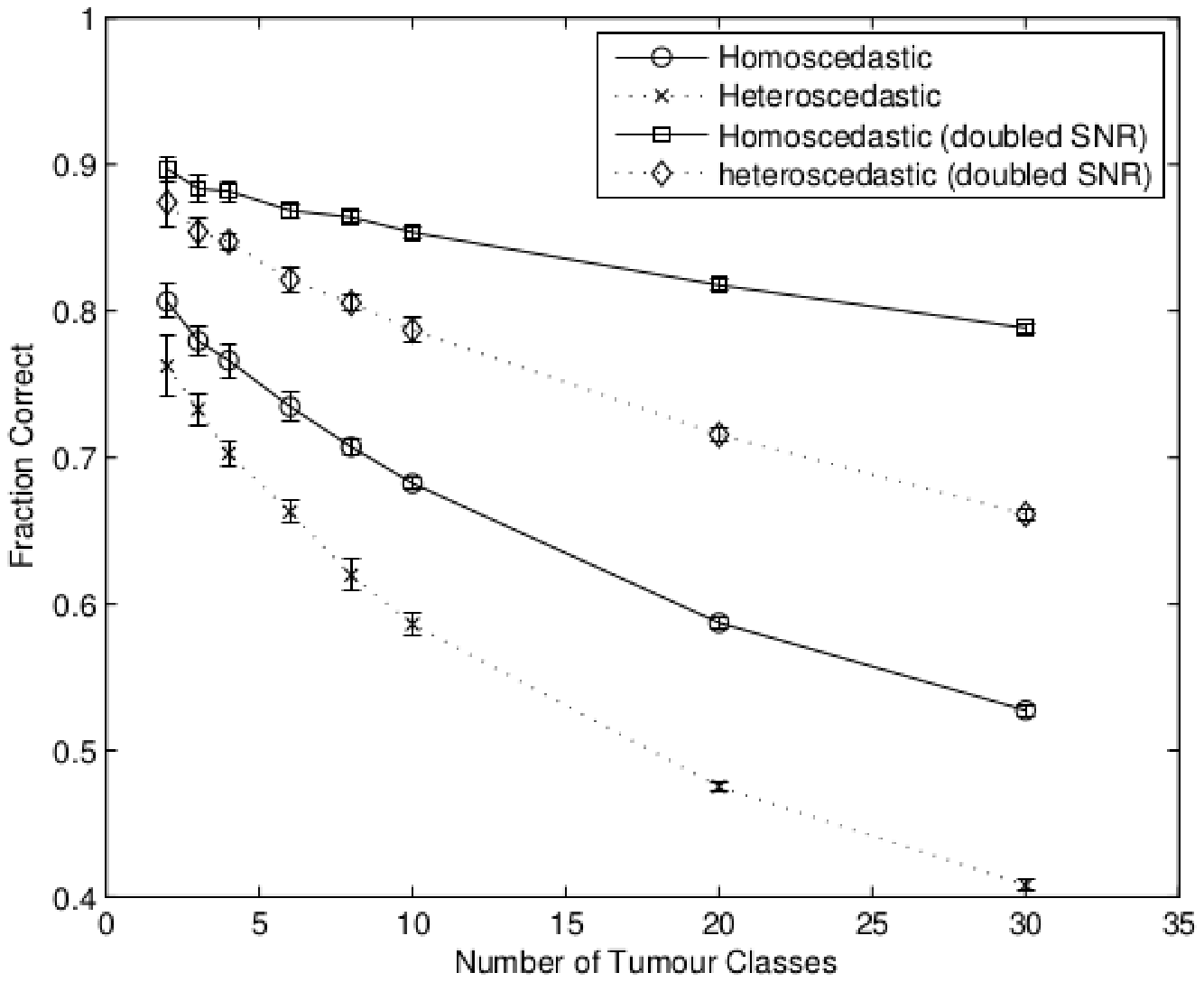}

		\caption{\it Results of the Monte-Carlo reliability analysis showing the fraction of correctly classified tumours with both the $D_{stan}$ (heteroscedastic) and $D_{tran}$ (homoscedastic) measures. Results are shown for both the measured repeatability noise levels and for a doubling of the signal-to-noise ratio.
			}
			\label{fig:reliability}
\end{figure}

Figure \ref{fig:reliability} shows the results of the reliability analysis. For the raw data with parameter dependent noise the average classification accuracy for two tumour classes is approximately 77\%. The classification accuracy decreases in a monotonic fashion as the classification density is increased. With 30 tumour classes the average accuracy is around 40\%. Results are also shown for the doubling of the SNR. Transformation of the variables to the homoscedastic space significantly improves the classification accuracy in both cases.

\section*{Discussion}

In this work we have presented a statistically motivated approach to comparing multi-dimensional intra-subject whole tumour measurements to inter-subject measurements from DCE-MRI. Hepatic metastatic colorectal cancer was used as a model to assess the ability of multi-parametric DCE-MRI data to support stratification of phenotypic whole tumour variation. The liver is a preferential site for metastases in colorectal cancer, multiple lesions are common and metastatic disease is a common target in therapeutic trials of novel agents. Genetic heterogeneity has been described between primary tumours and metastases (\cite{Jones2008}) and there is evidence that phenotypic and genetic heterogeneity also exists between metastatic lesions within the same patient. Goasguen and colleagues (\cite{Goasguen2009}) reported significant variation in treatment response in 64\% of tumour fragments derived from different metastases within a single patient. This was also associated with considerable inter-metastatic heterogeneity in levels of gene expression. This is supported by retrospective analysis of the CAIRO and CAIRO II trials (\cite{vanKessel2013}) which demonstrated mixed response to therapy in 36\% of patients with multiple metastases, associated with a decreased median survival of 23.7 months compared with 36 months in patients with homogeneous response. 

The growing evidence that significant biological variation exists within and between metastatic deposits implies that heterogeneity of tumour response to different therapies might be observed if discriminatory biomarkers can be developed. Repeated or multiple tissue biopsies are clearly impractical giving rise to an increasing need for alternate non-invasive approaches. Imaging biomarkers (IB) provide a potential solution offering unique advantages over soluble or tissue-based biomarkers. Ideally, IB could be used to identify biological / genetic variations to support enrichment of clinical trial data and provide predictive information to guide therapy. However there remain substantial technical problems associated with the use of IBs in this context. Identification of biological variability within tumours requires the calculation of reliable and robust IBs from each voxel in the tumour. In practice such pixel-by-pixel mapping is complicated by with significant uncertainties (errors) related to physiological and biological variation within the tissue. For example, the accuracy of measurements of blood volume varies systematically with the measured value (\cite{Li2000}) and the error models associated with many IB demonstrate similar but more complex behaviour (\cite{Li2004}).

A repeated measurements data set composed of 4 MR derived IB in hepatic liver colorectal metastases was collected from 29 subjects (102 tumours). The measurements have differing scales and dynamic ranges, but further to this, Bland-Altman plots show that the variables have parameter dependent noise characteristics. Consequently, use of the original variables will lead to large differences being identified inappropriately, due to the differences in noise
characteristics rather than biology. This invalidates the simple use of combined data in their raw state.

The overall variations seen in each DCI-MRI variable are most likely dependent on factors such as field strength, flip angle, tumour volumes etc.  In this work we use a repeated measures dataset where the MRI parameters such as flip angles, field strength etc are constant across the dataset; tumour size and AIF were controlled as much as possible. Due to these controls it is valid to seek an empirical transformation that considers the net error distributions, based on the observed error characteristics. We assume that the error distributions observed from the repeated measures dataset contain all sources of error but that any error characteristics dependent on experimental factors are approximately constant and independent.

Whilst a more sophisticated model could be constructed, based directly on all separate forms of perturbation, more data would be needed in order to estimate its parameters and any additional improvements in statistical efficiency may be negligible if the dominant trend has already been captured.

A maximum-likelihood optimisation of a power law transform was used to transforms variables to a space where the noise distribution is independent of the parameter value itself. Thus, the transformation of the variables has allowed them to be robustly combined in a single distance metric that operates over a Euclidean space. Any \textit{significant} changes in this distance may now be more accurately attributed to measurable changes.

The intra-subject results show a high level of similarity between tumours with some evidence of inter-tumoral heterogeneity within single subjects. This has implications when using multiple tumours from individual subjects to boost summary statistics that assume independence. Thus, in this work, results limited to 2 metastasis per subject are the most meaningful. With this limit enforced, the method is able to detect significant differences between the inter- and intra- groups, showing more heterogeneity between subjects than within. 

With a view to assessing the ability of the derived IB to quantify whole tumour heterogeneity, surrogate datasets with known ground truth were generated by Monte Carlo simulation. These datasets were used to investigate the reliability of the derived IB in terms of correctly classifying individual tumours to their known class at varying levels of classification resolution. The results of this analysis show that, for a reasonable level of classification resolution of around 20 tumour types, with the raw variables fused in the standard fashion, the derived IB are only capable of correct classification with an average rate of 50\%.

Combining the information from the variables as described in this work improves the classification results at all levels of classification density and is agreement with our hypothesis. (1) Homogeneous noise more closely matches the assumptions made in statistical tests based on analysis of variance. Summary statistics, such as mean and variance, generated from the analysis are now a truer reflection of the properties found in the data. (2) Furthermore,
by conforming better to these assumptions, the statistical efficiency of the tests is improved (\cite{Box1964}).

The Monte Carlo shows that a doubling of the SNR (\cite{Krokos2017}) for each variable significantly improves the classification rate to around 83\% for 20 tumour types and is above 85\% for 10 tumour types and less (1 in 10 error rate). In terms of individuals this would give a level of confidence that may be of practical value in clinical use, especially with regard to subjects with multiple metastatic tumours.

In conclusion, we have demonstrated the use of appropriate data transforms and combination of DCE-MRI derived parameters that ensures the credible interpretation of statistical differences. The characteristics of the transformed variables allow principled combination of data from multiple IB to characterise individual tumour deposits and produces a significant improvement in discrimination when compared to conventional approaches. With current levels of SNR in derived IB for hepatic metastatic tumours robust stratification/classification of tumours is not reliable. However, the work of \emph{Krokos,2017} has shown that with improved models and fitting procedures the doubling of the SNR is possible.

Further work in this area will investigate the validity of using standard corrections to account for the destabilising effects of haematocrit variation on DCE-MRI parameters (See appendix A).

\section*{Appendix A: Estimation of Contrast Agent Concentration from Signal Intensity}

For a sequence that spoils the transverse magnetisation and produces $T_1$ contrast, the signal intensity is given by

\begin{equation}
S=S_0 \frac{sin(\alpha)(1-e^{-\frac{TR}{T_1}})}{(1-e^{-\frac{-TR}{T_1}})cos(\alpha)}
\end{equation}

\noindent where $TR$ is the repetition time and $\alpha$ is the flip angle.

Gd induces  a  shift  in  the  bulk  magnetic susceptibility (BMS), the resonance frequency of the water protons. This phenomenon is caused by the local  variations  in  the  magnetic  field  due  to a Gd  inhomogeneous  distribution  within  the  vessel  and  in  particular  at  the  boundaries  of  the  tissues.  The  $T_2$ and  $T_2^*$ relaxation  times  are  shortened  and  the  
equivalent  sequences benefit from this  effect.  However, for  a  disrupted  blood brain barrier,  when  the  contrast agent is leaked in the extra-vascular extra-cellular space, from the vessels where there is a much larger water concentration, the BMS effect  is  reduced  and $T_1$ times are  the  dominant  effects.  In  that  case,  the  relationship  between  the  
relaxation  rate  $R_1=1/T_1$ and  the  contrast  agent  concentration  in  blood  $C_b$,  for  standard  contrast  agent  doses, is given by

\begin{equation}
R_1=R_{10}+r_1 C_b
\end{equation}

\noindent where  $r_1$ is  the  spin-lattice  relaxivity  constant  (i.e.  the  ability  of  the  contrast  agent  to  enhance  the 
detected signal according to the caused increase in the proton relaxation rate). For Gd it is assumed to be equal to the invitro  value  of  4.5$s^{-1}mM^{-1}$. $R_{10}$ is the relaxation rate in the absence of the contrast agent ($1/T_{10}$).

Pre- and post-contrast measurements allow for these equations to be solved to give the contrast agent blood concentration, $C_b$. A detailed explanation of this process is beyond the scope of this article. For a more detailed description of the method used please see (\cite{Jackson2005}).  The  concentration  of  the  contrast  agent  in  blood can  then  be  converted  to  the  concentration  in  plasma  by  taking  into  account  the  haematocrit (since the contrast agent is distributed in plasma)(\cite{Tofts2010}) giving

\begin{equation}
C_p=\frac{C_b}{1-Hct}
\end{equation}


\section*{Acknowledgements}

The authors would like to acknowledge  Dr Mark Saunders in the recruitment and referral of patients included in this study. The data used in this work was funded by an investigator-led research grant from F. Hoffmann-La Roche Ltd, the Manchester Experimental Cancer Medicine Centre.

\section*{Funding}

This work was funded by CRUK (Grant C8742/A18097). The funding source had no part in the collection, analysis or the interpretation of data, in the writing or the decision to publish this manuscript. 

\bibliographystyle{unsrt}
\bibliography{probDenBib}

\end{document}